\documentclass[12pt]{article}
\usepackage[top=1in, bottom=1in, left=1in, right=1in]{geometry}

\usepackage{cite}
\usepackage{amsmath,amssymb,amsfonts}
\usepackage{algorithmic}
\usepackage{graphicx}
\usepackage{textcomp}
\usepackage{xcolor}
\usepackage{adjustbox}
\usepackage{booktabs}
\usepackage{multirow}
\usepackage{subcaption}
\usepackage{pgfplots}
\usepackage[absolute]{textpos}
\usepackage{float}
\usepackage{fancyhdr}

\begin{document}

\vspace{-1.5em}
\begin{flushright}
\end{flushright}

\begin{center}
    \Large \textbf{Gradient Attention Map Based Verification of Deep Convolutional Neural Networks \\ with Application to X-ray Image Datasets}
    
    \vspace{1em}
    \normalsize
    Omid Halimi Milani\textsuperscript{1}, Amanda Nikho\textsuperscript{2}, Lauren Mills\textsuperscript{2}, Marouane Tliba\textsuperscript{3}, \\
    Ahmet Enis Cetin\textsuperscript{1}, Mohammed H. Elnagar\textsuperscript{2} \\
    
    \vspace{0.5em}
    \textsuperscript{1}Department of Electrical and Computer Engineering, University of Illinois Chicago, IL, USA \\
    \textsuperscript{2}Department of Orthodontics, College of Dentistry, University of Illinois Chicago, IL, USA \\
    \textsuperscript{3}University of Orleans, Orleans, France
    
\end{center}

\begin{abstract}

Deep learning models have great potential in medical imaging, including orthodontics and skeletal maturity assessment. However, applying a model to data different from its training set can lead to unreliable predictions that may impact patient care. To address this, we propose a comprehensive verification framework that evaluates model suitability through multiple complementary strategies. First, we introduce a Gradient Attention Map (GAM)-based approach that analyzes attention patterns using Grad-CAM and compares them via similarity metrics such as IoU, Dice Similarity, SSIM, Cosine Similarity, Pearson Correlation, KL Divergence, and Wasserstein Distance. Second, we extend verification to early convolutional feature maps, capturing structural misalignments missed by attention alone. Finally, we incorporate an additional garbage class into the classification model to explicitly reject out-of-distribution inputs. Experimental results demonstrate that these combined methods effectively identify unsuitable models and inputs, promoting safer and more reliable deployment of deep learning in medical imaging.

\end{abstract}

\section*{Keywords}
Attention Misalignment, Medical Imaging, Model Suitability, Explainable AI, Deep Learning Verification

\section{Introduction}

Deep learning has significantly improved automated diagnostic capabilities in medical imaging, particularly in orthodontics and maxillofacial surgery, where precise classification of conditions is essential for treatment planning. Orthodontic diagnosis tasks require high model accuracy. Ensuring model suitability for specific datasets remains a challenge. 

(i) If a model trained on one dataset is mistakenly tested on a completely different dataset, its decisions become unreliable.  
(ii) Transfer learning is widely used in practice. However, minor training after transfer learning may not be enough to adapt the model effectively to new datasets.  
(iii) An unrelated image may be applied to the neural network by mistake, and the model will generate a result based solely on its training data, leading to incorrect classifications. Additionally, X-ray scanners may capture only a portion of an image or scan a corrupted image, causing the model to produce misleading results without recognizing the input as incomplete or corrupted.  

Current evaluation methods primarily focus on performance metrics such as accuracy and confidence scores, which fail to indicate whether the model is inherently suitable for the test data. 

In orthodontics, precise classification is crucial for determining the appropriate treatment pathway. For example, in Class III malocclusion diagnosis, lateral cephalometric X-rays and profile photographs are commonly used to classify whether a patient requires surgical intervention \cite{Lisboa2017, Kawai2021}. Similarly, skeletal maturation assessment using cranial base synchondroses, such as the spheno-occipital synchondrosis (SOS), is vital for forensic age estimation and orthodontic treatment planning \cite{powell1963closure, al-gumaei2022comparison}. In both applications, misclassification can lead to incorrect diagnoses and suboptimal treatment decisions, emphasizing the need for reliable models.
Despite advancements in deep learning, model misapplication remains a critical issue. For example, if a model trained on malocclusion classification is mistakenly tested on SOS data, it may generate incorrect predictions due to fundamental differences in anatomical structures and classification objectives. Traditional evaluation metrics, such as accuracy, precision, and recall, do not provide insight into whether the model was trained on a compatible dataset.

To ensure consistency and clinical relevance, our model selection and experimental setup build on previously validated deep learning frameworks for orthodontic diagnosis and treatment planning. Specifically, we utilize models developed in prior work on third molar development classification \cite{milani2024fully} and treatment decision-making for anterior open bite cases \cite{rhee2025integrating}, both of which demonstrate robust performance on cephalometric, CBCT, and tabular data.



This problem necessitates a new model verification approach to ensure that a model is being used correctly. The key contributions of this work are as follows:

\begin{itemize}
    \item Introduce a Gradient Attention Map-based approach to verify whether a deep learning model is suitable for a given dataset or an image.
    \item Develop a feature map-based verification method that analyzes early convolutional activations to capture structural inconsistencies beyond attention patterns.
    \item Introduce a garbage class strategy to explicitly detect out-of-distribution or misapplied inputs, improving both model robustness and classification performance.
    \item Enhance model explainability by systematically evaluating model attention and feature patterns to ensure reliability in clinical deployment.
\end{itemize}

\section{X-ray Image Datasets}
\label{sec:dataset}

Deep learning models for medical imaging require datasets that are well-annotated and clinically validated to ensure reliability. This study utilizes two distinct datasets: one for spheno-occipital synchondrosis (SOS) fusion staging and another for Class III malocclusion classification using lateral cephalometric X-rays. Both datasets are carefully curated to support automated classification tasks while addressing the challenge of model verification.

\subsection{SOS Fusion Staging Dataset}

This dataset consists of cone-beam computed tomography (CBCT) scans collected from patients with varying stages of spheno-occipital synchondrosis fusion. 


To standardize the dataset, each CBCT scan was preprocessed. The region of interest (ROI) was extracted through manual segmentation by expert, aligning the planes to ensure consistency.


\subsection{Class III Malocclusion X-ray Dataset}


This dataset consists of lateral cephalometric X-ray images: orthodontic camouflage and surgical intervention. 



Each patient record includes X-ray images paired with expert annotations. 

The dataset was curated to include only cases with confirmed successful treatment outcomes, ensuring that the classification labels reflect reliable clinical decisions.





\section{Methodology}

After curating and preprocessing the dataset for consistency and reliability, we utilize advanced deep learning models for automated classification. Pre-trained models known for their robustness in image analysis are employed to enhance classification accuracy while reducing the need for extensive new model training. The selected models include ResNet18, ResNet34, ResNet50, ConvNeXt, and ConvNeXt with attention mechanisms. These models have been fully trained and have demonstrated strong performance in classification tasks \cite{liu2022convnet, he2016deep, koonce2021resnet, koonce2021resnet}.



\subsection{Model Verification Using Grad-CAM Similarity Metrics}

To quantify the alignment of candidate models with the reference model, we compute statistical similarity metrics between their Grad-CAM attention maps. These metrics capture spatial, structural, and distributional differences, enabling a robust evaluation of attention alignment \cite{livieris2023explainable,selvaraju2017grad, lecun1998}.
\subsubsection{Grad-CAM Computation}
For each input image $I$, the Grad-CAM heatmap $H_M(I)$ for a given model $M$ is computed as:
\begin{equation}
H_M(I) = \operatorname{ReLU} \left( \sum_{k} \alpha_k A_k \right)
\end{equation}

\subsubsection{Threshold-Based Binarization}
To obtain a fair comparison, we convert raw Grad-CAM heatmaps into binary masks using a median thresholding approach \cite{doshi2017towards}:
\begin{equation}
B_{M}(i, j) = 
\begin{cases} 
1, & H_{M}(i, j) > \operatorname{median}(H_{M}) \\
0, & \text{otherwise}
\end{cases}
\end{equation}
where $B_{M}(i, j)$ represents the binarized attention region for model $M$.

\subsubsection{Spatial Overlap Metrics}
To measure the degree of similarity between the reference and candidate model’s focus regions, we compute the Intersection over Union (IoU) and Dice Similarity Coefficient (DSC) \cite{adebayo2018sanity,livieris2023explainable,pellano2024movements}:
\begin{equation}
\operatorname{IoU}(M_1, M_2) = \frac{|B_{M_1} \cap B_{M_2}|}{|B_{M_1} \cup B_{M_2}|}
\end{equation}
\begin{equation}
\operatorname{DSC}(M_1, M_2) = \frac{2 |B_{M_1} \cap B_{M_2}|}{|B_{M_1}| + |B_{M_2}|}
\end{equation}

\subsubsection{Perceptual Similarity Assessment}
To further analyze attention alignment, we compute the Structural Similarity Index (SSIM), which captures perceptual differences in intensity distribution:
\begin{equation}
\operatorname{SSIM}(M_1, M_2) = \frac{(2 \mu_{M_1} \mu_{M_2} + C_1)(2 \sigma_{M_1M_2} + C_2)}{(\mu_{M_1}^2 + \mu_{M_2}^2 + C_1)(\sigma_{M_1}^2 + \sigma_{M_2}^2 + C_2)}
\end{equation}

\cite{samek2019explainable,selvaraju2020grad,hase2020evaluating}.

\subsubsection{Statistical Correlation Analysis}
To measure linear alignment in pixel intensities, we compute Cosine Similarity and Pearson Correlation \cite{ras2022explainable,li2021experimental}:
\begin{equation}
\operatorname{Cosine}(M_1, M_2) = \frac{H_{M_1} \cdot H_{M_2}}{\|H_{M_1}\| \|H_{M_2}\|}
\end{equation}
\begin{equation}
r(M_1, M_2) = \frac{\sum (H_{M_1} - \bar{H}_{M_1}) (H_{M_2} - \bar{H}_{M_2})}{\sqrt{\sum (H_{M_1} - \bar{H}_{M_1})^2} \sqrt{\sum (H_{M_2} - \bar{H}_{M_2})^2}}
\end{equation}

\subsubsection{Distributional Divergence Measures}
To assess differences in the probability distributions of attention maps, we compute Kullback-Leibler (KL) Divergence and Wasserstein Distance \cite{chaddad2023survey, chakraborty2022generalizing,panaretos2019statistical}:
\begin{equation}
D_{KL}(P_{M_1} \| P_{M_2}) = \sum_{i} P_{M_1}(i) \log \frac{P_{M_1}(i)}{P_{M_2}(i) + \epsilon}
\end{equation}
\begin{equation}
W(P_{M_1}, P_{M_2}) = \sum_{i} d(i) |P_{M_1}(i) - P_{M_2}(i)|
\end{equation}


\subsubsection{Integrating Similarity Metrics into the Verification Framework}
The computed similarity metrics serve as features for a Random Forest classifier, enabling automated differentiation between well-aligned and misaligned models. By learning from these statistical patterns, the framework systematically identifies models whose attention maps exhibit clinically meaningful alignment with expert-defined regions, ensuring more reliable deep learning-based decision-making.


\begin{figure}[htbp]
    \centering
    \includegraphics[width=0.48\textwidth]{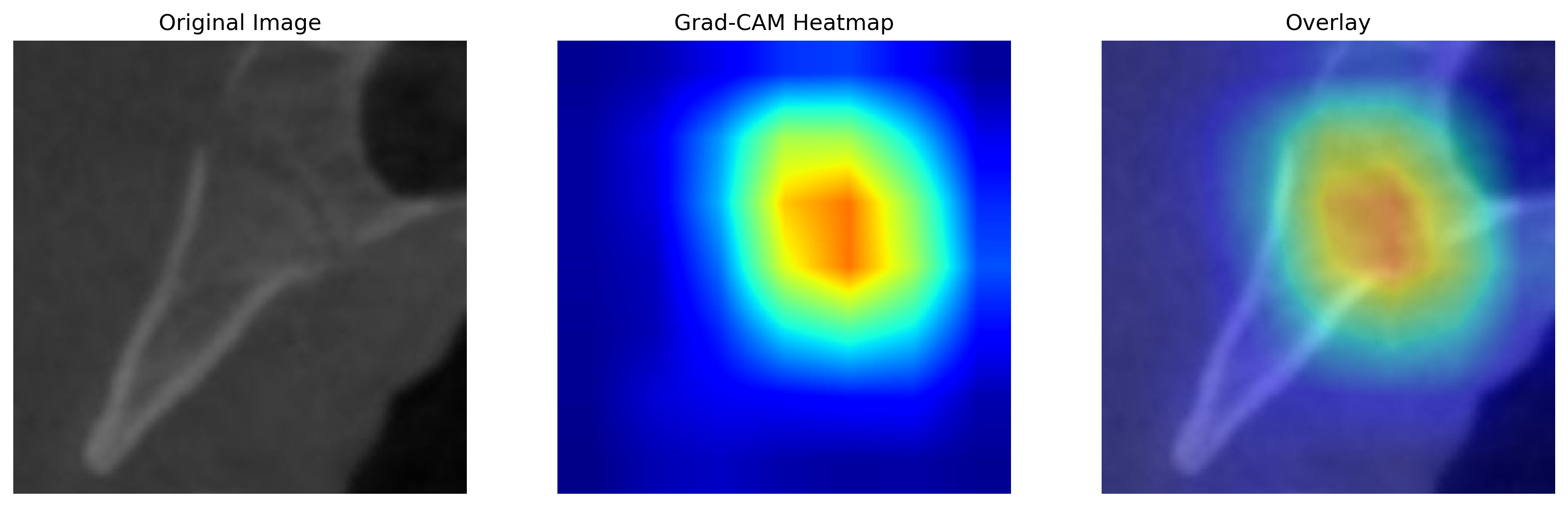}
    \caption{Grad-CAM visualization for a model trained and tested on the SOS dataset. The attention is well-aligned with the anatomical structures, highlighting relevant areas.}
    \label{fig:gradcam_correct}
\end{figure}

\begin{figure*}[htbp] 
    \centering 
    \includegraphics[width=\textwidth]{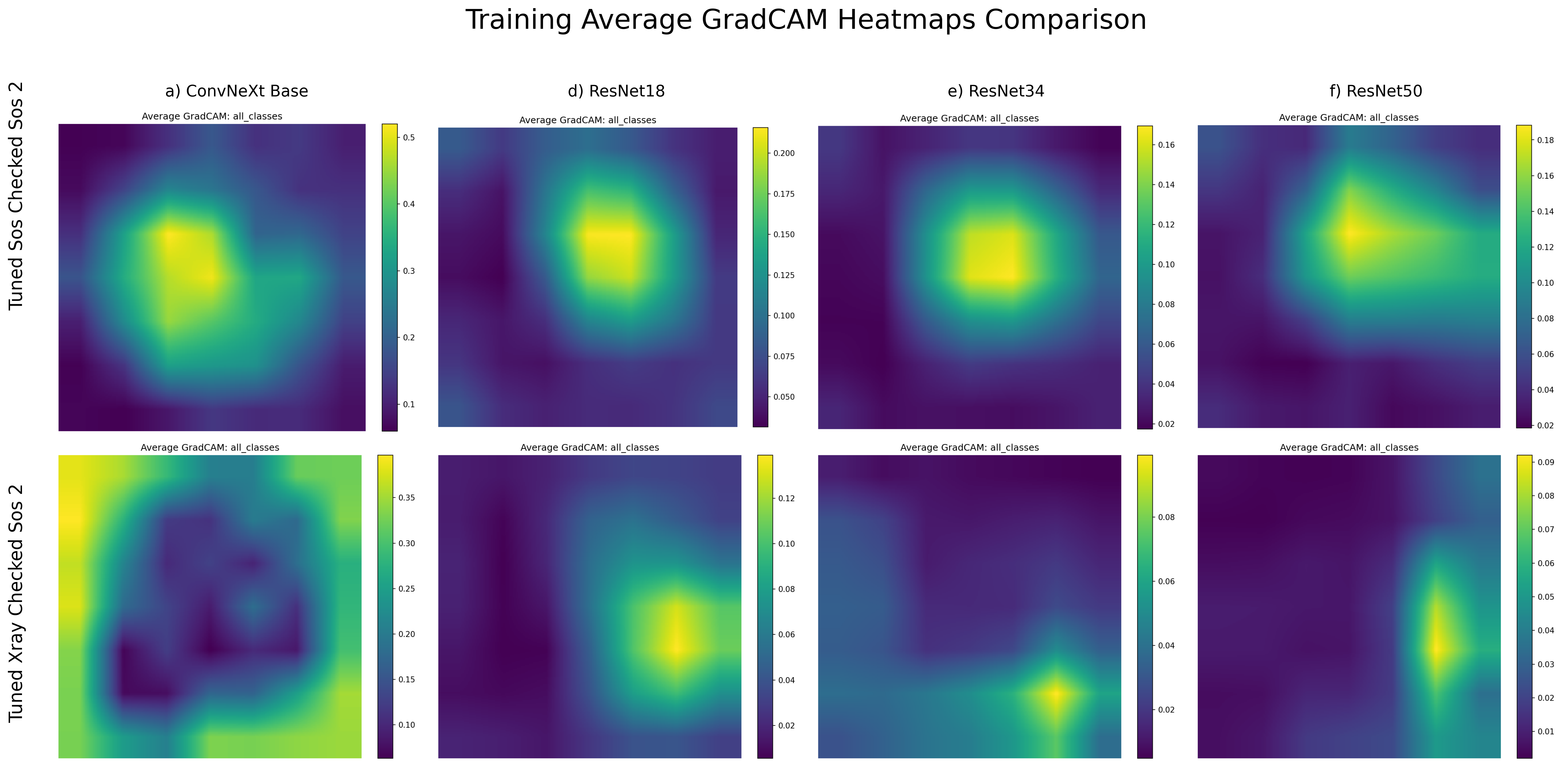} 
    \caption{Visualization of the average Gradient Attention Maps (GAMs) for each model during training. 
    The first row represents attention maps for models trained on the SOS dataset and evaluated on the SOS training set. 
    The second row shows attention maps for models trained on the Class III dataset and tested on the SOS training set. The heatmaps illustrate where each model focuses when making predictions.
    }
    \label{fig:train_vs_test_gam} 
\end{figure*}
\subsection{Random Forest-based Model Verification and Rejection Strategy}

Traditional evaluation metrics, such as classification accuracy, fail to indicate whether a neural network model is genuinely suitable for a specific classification task. A model may achieve high accuracy during training while relying on spurious correlations or focusing on irrelevant anatomical features, leading to unreliable decision-making in real-world clinical applications. To address this challenge, we introduce a random forest-based  model verification framework using interpretable learning-based classification.

Our approach trains a classification model to distinguish between clinically meaningful and misaligned attention patterns. The Random Forest classifier learns patterns in attention alignment by leveraging statistical representations of Grad-CAM differences, distinguishing valid models from those with inconsistent or irrelevant focus regions using the features described in the previous section.

This method improves model selection by quantifying how well a model mimics expert decision-making, contributing to a more transparent and explainable AI-driven clinical workflow. By systematically rejecting models with misaligned focus regions, we ensure that only models that correctly align their attention with expert-defined regions are considered suitable.



The effectiveness of this approach is demonstrated in Fig.~\ref{fig:gradcam_correct}, which shows a Grad-CAM visualization for a model trained and tested on the SOS dataset. The model correctly attends to relevant anatomical structures, highlighting the importance of attention alignment as an evaluation metric beyond traditional performance scores.



\subsection{Feature Map-Based Verification via Internal Representations (Method 2)}

To further enhance model verification, we extract internal feature maps from early convolutional layers of the network. These feature maps often contain edge-like or low-level patterns that can indicate whether the model is processing the input in a clinically meaningful manner.

Let $f_i(k,l)$ denote the activation value at location $(k,l)$ from the $i$-th filter of a selected convolutional layer. We define the overall feature response at location $(k,l)$ as:
\begin{equation}
S(k,l) = \sum_{i} |f_i(k,l)|
\end{equation}

Since ReLU activation ensures non-negativity, $S(k,l)$ can be treated as a 2D distribution. We compute these maps for both the candidate and reference models across Layer 1 and Layer 2. Then, the same suite of similarity metrics (IoU, Dice, SSIM, Cosine, Pearson, KL, and Wasserstein) is used to compare these distributions.

This method is particularly effective in detecting models trained on unrelated datasets or misaligned input data. Random forest classifiers trained on these similarity features can distinguish between suitable and unsuitable models, even when classification accuracy appears acceptable.

\subsection{Garbage Class Strategy for Rejecting Out-of-Distribution Inputs (Method 3)}

We also introduce an auxiliary strategy based on extending the classification space. Instead of designing binary or multi-class classifiers alone, we define a new $k+1$-class classification problem where the additional class represents "non-target" or out-of-distribution samples.

For instance, in the case of 5-stage SOS classification, we define a 6-class network where the sixth class corresponds to inputs that do not belong to any SOS category (e.g., Chin or CVM images). This garbage class is trained using irrelevant but structurally similar X-ray images.

During inference, the model can effectively reject unfamiliar inputs by assigning them to the $k+1$-th class. This technique acts as a built-in verification mechanism and prevents the network from forcing a prediction on unsuitable data.

Our experiments show that models with this extended class formulation achieve better robustness and produce highly accurate classifications for suitable inputs, while flagging mismatched ones for rejection.

\section{Results and Evaluation}
To demonstrate our verification framework, we visualize average Gradient Attention Maps (GAMs) from models trained and tested on the SOS dataset versus those trained on Class III data and tested on SOS. As shown in Figure~\ref{fig:train_vs_test_gam}, SOS-trained models focus on relevant anatomical regions, while Class III-trained models exhibit misaligned attention. 

Table~\ref{tab:combined_results} presents the performance comparison of selected deep learning models on two datasets: CLASS III (X-ray) and the SOS dataset. The evaluation metrics include Accuracy, Precision, Recall, and F1-Score. The results indicate that ConvNeXt outperforms other models in both datasets, achieving the highest scores across all metrics in CLASS III and showing competitive performance on the SOS dataset. The reference model, ConvNeXt+Attn, is only evaluated on the SOS dataset, where it achieves the best overall performance, surpassing all other models. These results highlight the effectiveness of attention-based architectures in improving model generalization and robustness across different datasets.

\begin{table}[t!]
\centering
\caption{Performance Comparison of Selected Models on CLASS III (Chin) and SOS Datasets}
\label{tab:combined_results}
\resizebox{\linewidth}{!}{%
\begin{tabular}{lcccc|cccc}
\toprule
\textbf{Model} & \multicolumn{4}{c}{\textbf{CLASS III - Chin}} & \multicolumn{4}{c}{\textbf{SOS Dataset}} \\
\cmidrule(lr){2-5} \cmidrule(lr){6-9}
 & \textbf{Accuracy} & \textbf{Precision} & \textbf{Recall} & \textbf{F1-Score} & \textbf{Accuracy} & \textbf{Precision} & \textbf{Recall} & \textbf{F1-Score} \\
\midrule
ResNet-18  & 86.36 & 88.60 & 90.18 & 89.38 & 72.55 & 72.12 & 72.55 & 70.84 \\
ResNet-34  & 87.50 & 88.79 & 91.96 & 90.35 & 75.07 & 75.01 & 75.07 & 74.05 \\
ResNet-50  & 85.80 & 92.23 & 84.82 & 88.37 & 74.79 & 74.08 & 74.79 & 73.79 \\
ConvNeXt   & \textbf{90.91} & \textbf{92.11} & \textbf{93.75} & \textbf{92.92} & 78.99 & 78.57 & 78.99 & 78.35 \\
ConvNeXt+Attn (reference model) & - & - & - & - & \textbf{79.97} & \textbf{80.65} & \textbf{79.97} & \textbf{78.87} \\
\bottomrule
\end{tabular}%
}
\end{table}

\begin{table}[t!]
\centering
\caption{Performance of the Random Forest Model}
\label{tab:rf_results}
\resizebox{\linewidth}{!}{%
\begin{tabular}{lccccc}
\toprule
\textbf{Model} & \textbf{Accuracy (\%)} & \textbf{Precision (\%)} & \textbf{Recall (\%)} & \textbf{F1-Score (\%)} & \textbf{ROC AUC (\%)} \\
\midrule
Random Forest  & 87.54 & 87.54 & 87.54 & 87.55 & 93.52 \\
\bottomrule
\end{tabular}%
}
\end{table}

\begin{table}[htbp]
\centering
\caption{Validation Statistics of Grad-CAM Similarity Metrics for Misaligned (Label 0) and Aligned (Label 1) Models}
\label{tab:gradcam_stats}
\resizebox{\linewidth}{!}{%
\begin{tabular}{lccccc}
\toprule
\textbf{Metric} & \textbf{Label} & \textbf{Mean} & \textbf{Min} & \textbf{Max} & \textbf{Std} \\
\midrule
IoU & 0 & 0.3168 & 0.0000 & 0.7930 & 0.1824 \\
Dice & 0 & 0.4516 & 0.0000 & 0.8845 & 0.2181 \\
SSIM & 0 & 0.1980 & 0.0233 & 0.7011 & 0.1167 \\
Cosine & 0 & 0.3887 & 0.0261 & 0.8755 & 0.2219 \\
Pearson & 0 & -0.0193 & -0.7909 & 0.8063 & 0.4029 \\
KL Divergence & 0 & 7.1166 & 0.2194 & 17.7622 & 4.7189 \\
Wasserstein & 0 & 0.1714 & 0.0681 & 0.2792 & 0.0350 \\
\midrule
IoU & 1 & 0.4558 & 0.0828 & 0.7770 & 0.1492 \\
Dice & 1 & 0.6116 & 0.1529 & 0.8745 & 0.1443 \\
SSIM & 1 & 0.3845 & 0.0902 & 0.7038 & 0.1309 \\
Cosine & 1 & 0.6952 & 0.1561 & 0.9559 & 0.1749 \\
Pearson & 1 & 0.4037 & -0.7668 & 0.9403 & 0.3782 \\
KL Divergence & 1 & 3.2797 & 0.1644 & 12.2506 & 2.4379 \\
Wasserstein & 1 & 0.1297 & 0.0602 & 0.2010 & 0.0358 \\
\bottomrule
\end{tabular}
}
\end{table}


The effectiveness of our verification approach is further reinforced by Figure~\ref{fig:train_vs_test_gam}, which consolidates the GAM comparisons across models. This demonstrates how our method systematically removes models that focus on incorrect features while retaining those with aligned attention patterns, ensuring that models are not only performing well in terms of accuracy but also making clinically meaningful decisions.

\subsection{Model Verification Outcomes}

The proposed model verification framework effectively identifies models that align with the expected learning patterns by leveraging interpretable decision models trained on features extracted from Gradient Attention Maps (GAMs). This approach ensures that only models optimized for the SOS dataset are retained, preventing the inclusion of models trained on unrelated data that could compromise clinical decision-making.

Instead of relying on predefined separation values, we employ a random forest classifier trained to recognize patterns in how models attend to different anatomical regions. By extracting statistical representations of GAM differences between a reference model and candidate models, the classifier learns distinguishing characteristics that differentiate well-aligned models from those that focus on irrelevant or inconsistent regions.

Our evaluation demonstrates that this method successfully filters out models that do not exhibit alignment with clinically meaningful features. Even among models trained on the SOS dataset, those showing inconsistent GAM patterns—indicating attention to incorrect anatomical structures—were effectively excluded. This underscores the importance of attention-based model validation, ensuring that deep learning models focus on relevant medical regions rather than just maximizing traditional performance metrics.

For evaluation, we compute the gradient attention map for each sample and compare it against the reference average gradient attention map. From this comparison, we extract seven similarity features that quantify spatial, perceptual, and statistical similarities between the attention maps. This results in a dataset of 586 data points, where each sample is characterized by these seven features.

Among these data points, 312 are classified as \textit{not acceptable}, meaning they were tested on a model that was not trained on similar data, while 274 are categorized as \textit{acceptable}, indicating they were tested on a model trained on similar data.

To assess the model’s performance, we apply 5-fold cross-validation using a random forest classifier. The model achieves a Receiver Operating Characteristic - Area Under the Curve (ROC AUC) score of 93.52\%, with an overall accuracy of 87.54\%. The precision, recall, and F1-score are consistently around 87.54\%, with an average F1-score of 87.55\% across all samples.



By leveraging an interpretable learning-based approach, we ensure that models exhibiting attention inconsistencies are systematically identified and removed.



\subsection{Feature Map-Based Verification Results (Method 2)}
We further evaluate the model’s suitability using internal representations extracted from the first and second convolutional layers. These feature maps are transformed into probability distributions and compared using the same seven similarity metrics as in Method 1. A Random Forest classifier is trained to distinguish between suitable and unsuitable models.

Table~\ref{tab:method2} shows that using Layer 1 activations from a model trained on the Chin dataset and tested on SOS samples results in near-perfect performance, with 99.36\% accuracy and 99.96\% ROC AUC. The performance slightly drops when testing against rotated SOS images, yet it remains robust above 79\%.

\begin{table}[H]
\centering
\caption{Rotation Robustness Comparison using Feature Map-Based Verification (Method 2)}
\label{tab:method2}
\resizebox{\linewidth}{!}{%
\begin{tabular}{lllccc}
\toprule
\textbf{Model} & \textbf{Train} & \textbf{Test} & \textbf{Accuracy} & \textbf{ROC AUC} & \textbf{F1 (avg)} \\
\midrule
Layer 1        & Chin           & SOS           & 0.9936 & 0.9996 & 0.9936 \\
Gradient-Based & Chin           & SOS           & 0.8754 & 0.9358 & 0.8749 \\
Layer 1        & SOS            & Rotated SOS   & 0.8013 & 0.8825 & 0.8005 \\
Layer 2        & SOS            & Rotated SOS   & 0.7917 & 0.8690 & 0.7912 \\
\bottomrule
\end{tabular}
}
\end{table}
\subsection{Garbage Class Strategy Results (Method 3)}
To detect out-of-distribution inputs, we introduce a $k+1$ class for models originally trained for $k$-class classification. This additional "garbage" class is trained using unrelated datasets such as Chin or CVM samples.

Tables~\ref{tab:chin6class} and \ref{tab:cvm6class} demonstrate that the 6th class achieves perfect 1.00 precision, recall, and F1-score, indicating its ability to correctly isolate mismatched data. Moreover, overall model accuracy increases when using this approach, showing not only its rejection power but also enhanced performance.

\vspace{0.8em}
\textbf{With Chin as the 6th class:}
\begin{table}[H]
\centering
\caption{Classification with Chin as Garbage Class (Method 3)}
\label{tab:chin6class}
\resizebox{\linewidth}{!}{%
\begin{tabular}{lcccc}
\toprule
\textbf{Class} & \textbf{Precision} & \textbf{Recall} & \textbf{F1-Score} & \textbf{Support} \\
\midrule
1 & 0.92 & 0.92 & 0.92 & 159 \\
2 & 0.78 & 0.73 & 0.75 & 92 \\
3 & 0.71 & 0.82 & 0.76 & 92 \\
4 & 0.70 & 0.62 & 0.66 & 125 \\
5 & 0.88 & 0.89 & 0.89 & 255 \\
\textbf{6} & \textbf{1.00} & \textbf{1.00} & \textbf{1.00} & \textbf{176} \\
\midrule
\textbf{Avg} & 0.83 & 0.83 & 0.83 & 899 \\
\end{tabular}
}
\end{table}

\vspace{0.5em}
\textbf{With CVM as the 6th class:}
\begin{table}[H]
\centering
\caption{Classification with CVM as Garbage Class (Method 3)}
\label{tab:cvm6class}
\resizebox{\linewidth}{!}{%
\begin{tabular}{lcccc}
\toprule
\textbf{Class} & \textbf{Precision} & \textbf{Recall} & \textbf{F1-Score} & \textbf{Support} \\
\midrule
1 & 0.86 & 0.91 & 0.88 & 159 \\
2 & 0.69 & 0.68 & 0.69 & 92 \\
3 & 0.72 & 0.74 & 0.73 & 92 \\
4 & 0.70 & 0.57 & 0.63 & 125 \\
5 & 0.87 & 0.91 & 0.89 & 255 \\
\textbf{6} & \textbf{1.00} & \textbf{1.00} & \textbf{1.00} & \textbf{181} \\
\midrule
\textbf{Avg} & 0.81 & 0.80 & 0.80 & 904 \\
\end{tabular}
}
\end{table}

\section{Conclusion}
\label{sec:conclusion}
We proposed a verification framework that assesses the suitability of deep learning models in medical imaging by analyzing Gradient Attention Maps (GAM). Beyond Grad-CAM-based verification, we introduced a feature map-based method using early convolutional activations, and a garbage class strategy to reject out-of-distribution inputs. Together, these approaches enhance model reliability and prevent misapplication. Future work will focus on extending the framework to multi-modal datasets.

\section*{Acknowledgment}
This material is based upon work supported by the National Science Foundation under Award No. 2303700.

\end{document}